\documentclass[preprint,amsmath,amssymb,aps]{revtex4-2}

\usepackage{graphicx}
\usepackage{dcolumn}
\usepackage{bm}
\usepackage{physics}
\usepackage{filecontents}
\usepackage{xcolor}

\begin{document}

\title{Free scalar correlators in de Sitter via the stochastic approach beyond slow roll}

\author{Archie Cable}
 \email{archie.cable18@imperial.ac.uk}
\author{Arttu Rajantie}%
 \email{a.rajantie@imperial.ac.uk}
\affiliation{Department of Physics,\\Imperial College London,\\London, SW7 2AZ, United Kingdom}

\date{\today}

\maketitle

\textbf{Abstract}\\
    The stochastic approach to calculating scalar correlation functions in de Sitter spacetime is extended beyond the overdamped "slow roll" approximation. We show that with the correct noise term, it reproduces the exact asymptotic long-distance behaviour of field correlators in free field theory, thereby demonstrating the viability of the technique. However, we also show that the naïve way of calculating the noise term by introducing a cut-off at the horizon does not give the correct answer unless the cut-off is chosen specifically to give the required result. We discuss the implications of this for interacting theories.

\tableofcontents

\section{Introduction}
\label{sec:intro}

The study of spectator scalar fields in de Sitter via quantum field theory (QFT) in a curved spacetime is a well studied area \cite{Birrell-Davies_book,bunch_davies_1978,tagirov_QFT_deSitter,chernikov...de_Sitter_QFT}. Beyond formal interest, spectator scalars are applicable to many areas of cosmology, in particular the inflationary epoch \cite{Starobinsky:1980,guth_inflation,linde_inflation,slow_roll_liddle,Vazquez_Gonzalez_2020_slowroll}. Examples of this include the study of post-inflationary dynamics \cite{Hardwick_2017_post-inflation}, the generation of dark matter \cite{Peebles_1999_darkmatter, Hu_2000_darkmatter, Markkanen_2018_darkmatter}, the anisotropy of the gravitational wave background \cite{gwb_anisotropy} and the triggering of electroweak vacuum decay \cite{Espinosa_2008_vdecay,Herranen_2014_vdecay,Markkanen_2018_vdecay}.\\

Standard QFT techniques can be applied to free field theory in de Sitter; however, for theories including self-interactions, infrared divergences result in the breakdown of perturbation theory for light scalar fields \cite{allen_folacci_perturbative_corr, allen_perturbative_corr, Sasaki:1993, Suzuki:1994}, which has led physicists to explore other methods \cite{hu_oconner_symm_behaviour,boyanovsky_quantum_correct_SR,Serreau_2011,Herranen_2014,Gautier_2013,Gautier_2015,Arai_2012,Guilleux_2015}. The stochastic approach \cite{starobinsky_1986,Starobinsky-Yokoyama_1994} is one such method, where quantum behaviour can be approximated by a stochastic contribution to the classical equations of motion. It is becoming increasingly popular and is now widely used in the literature as an alternative to perturbative QFT \cite{ Tsamis_2005,Finelli_2009,Finelli_2010,Vennin_2015,Rigopoulos_2016,Moss_2017,Hardwick_2017_post-inflation,Tokuda_2018,Tokuda_2018_2,Glavan_2018,Cruces_2019,Grain_2017,Firouzjahi_2019,Markkanen_2019,Pinol_2019,Hardwick_2019,Pattison_2019,Markkanen_2020,Moreau_2020,Moreau:2020gib}. It is possible to derive the stochastic equations via a path integral, which aligns more with standard thermal field theory approaches \cite{Rigopoulos_2016,Moss_2017,Prokopec_2018,bounakis2020feynman,Garbrecht...2014_correlations_deSitter,Garbrecht_2015_Fdiag,morikawa_1990,rigopoulos2013fluctuationdissipation,Levasseur_2013, Pinol_2019, Pinol:2020}; however, in this paper we will focus on the original mode expansion method \cite{starobinsky_1986, Starobinsky-Yokoyama_1994}. One of the major attractions of the stochastic approach is that the scalar correlation functions can be considered as statistical quantities governed by the probability distribution function (PDF), leading to straightforward computations via a spectral expansion \cite{Markkanen_2019}.\\

In much of the current literature \cite{Cruces_2019,Glavan_2018,Vennin_2015,Grain_2017,Firouzjahi_2019,Markkanen_2019, Markkanen_2020,Moreau_2020,Moreau:2020gib}, the stochastic approach is applied to almost-massless scalar fields where one can use the overdamped ``slow roll'' approximation in which the second derivative of the field is negligible, reducing the equation of motion to a first-order differential equation. In this limit, the pioneering work of Starobinsky and Yokoyama \cite{starobinsky_1986, Starobinsky-Yokoyama_1994} showed that the existence of the de Sitter horizon allows one to split the modes between sub- and superhorizon wavelengths. The quantum subhorizon modes give rise to a stochastic noise term, which can be calculated by explicitly integrating out the modes with momenta above the horizon-scale cutoff. The long-time or long-distance asymptotic behaviour of field correlation functions obtained by solving the resulting stochastic equation agrees with the original quantum field theory result, demonstrating the validity of the approach.\\

However, there are many situations in which the slow roll approximation is not valid, because the scalar field mass $m$ is not much smaller than the Hubble rate $H$. As long as $m<\sqrt{2}H$, the long-wavelength modes are still amplified by the expansion of space, and at asymptotically long distances the stochastic approach should still be valid.
The extension of the stochastic approach to these cases, beyond the slow-roll limit, has been discussed in Refs. \cite{Grain_2017,Cruces_2019,Pattison_2019} in the top-down picture by deriving the effective stochastic theory from the microscopic quantum field theory description using a similar cutoff procedure as in the slow-roll case. In this paper, we consider the same question from the bottom-up perspective, asking whether and when the stochastic theory actually reproduces the correct quantum field theory behaviour in a free field theory where the latter is exactly known.\\

We find that this is not the case for the stochastic theory obtained with the cut-off approach: the asymptotic behaviour of the correlators does not agree with the quantum field theory results. On the other hand, with a different mass-dependent noise term it is possible to reproduce all equal-time QFT correlators correctly. This gives confidence that the stochastic approach itself can be valid beyond the slow-roll limit, but it leaves open the question of how to compute the correct noise term in general. In the free field case, we can determine it by matching the correlators, but in an interacting theory that is not possible because no exact QFT result is known. Nonetheless, this current work with free fields paves the way for future work on interacting theories.\\

The paper is organised as follows. In Section \ref{sec:stochastic_approach}, we begin with a brief overview of Starobinsky and Yokoyama's original approach before extending it beyond the overdamped limit. In Section \ref{sec:qft_comparison}, we solve the stochastic system and compare it with the full quantum field theory.
Finally, we discuss our findings and conclusions in Section \ref{sec:discussion}.\\

\section{The stochastic approach}
\label{sec:stochastic_approach}

\subsection{The overdamped limit}
\label{subsec:1PDF}

Let us consider a scalar quantum field $\hat{\phi}(t,\mathbf{x})$ in a cosmological de Sitter spacetime with scale factor $a(t)=e^{Ht}$. We define the effective mass $m^2=m_0^2+12\xi H^2$, where $m_0$ is the scalar mass and $\xi$ is the non-minimal coupling constant\footnote{Later, we introduce $\xi(t,\mathbf{x})$ to represent a stochastic noise. This is not to be confused with the $\xi$ used here to represent the non-minimal coupling constant.}, such that the scalar potential $V(\hat{\phi})$ incorporates all non-minimal interactions. We introduce the field derivative $\hat{\pi}=\Dot{\hat{\phi}}$ as an independent variable such that the equations of motion are written as\\ 

\begin{equation}
    \label{2PDE_eom}
    \begin{pmatrix}\Dot{\hat{\phi}}\\\Dot{\hat{\pi}}\end{pmatrix}=\begin{pmatrix}\hat{\pi}\\-3H\hat{\pi}+\frac{1}{a(t)^2}\nabla^2\hat{\phi}-V'(\hat{\phi})\end{pmatrix},
\end{equation}\\

\noindent where dots and primes denote derivatives with respect to time and field respectively, and $H=\Dot{a}/a$ is the Hubble rate which we take to be constant.\\

The original stochastic approach proposed by Starobinsky and Yokoyama \cite{starobinsky_1986, Starobinsky-Yokoyama_1994} splits the field into classical and quantum components\\

\begin{equation}
    \label{field_split}
    \hat{\phi}=\overline{\phi}+\hat{\delta\phi}
\end{equation}\\

\noindent about the de Sitter horizon scale $1/H$, where $\overline{\phi}$ contains the classical superhorizon ($k/a(t)<1/H$) modes and\\

\begin{equation}
    \label{subhorizon_field}
    \hat{\delta\phi}(t,\mathbf{x})=\int\frac{d^3\mathbf{k}}{(2\pi)^3}W_k(t)\hat{\phi}_k(t,\mathbf{x})
\end{equation}\\

\noindent contains the quantum subhorizon ($k/a(t)>1/H$) modes. The field operator can be written as $\hat{\phi}_k(t,\mathbf{x})=\phi_k(t,\mathbf{x})\hat{a}_k+\phi^*_k(t,\mathbf{x})\hat{a}_k^{\dagger}$, where $\{\hat{a}_k\}$ are the set of creation and annihilation operators relating to the Bunch-Davies vacuum \cite{bunch_davies_1978,tagirov_QFT_deSitter,Birrell-Davies_book}, and the window function $W_k(t)=\theta(k-\epsilon a(t)H)$ acts as a cut-off to separate the subhorizon quantum modes from the superhorizon classical ones, where the real parameter $\epsilon$ determines the precise cut-off scale.\\

In the overdamped ``slow roll'' limit, where $V''(\hat{\phi}) \ll H^2$, which corresponds to (nearly) massless scalar fields,
$\Dot{\hat{\pi}}$ is negligible. Therefore, one obtains a 1-dimensional equation of motion\\

\begin{equation}
    \label{overdamped_eom}
    0=\Dot{\hat{\phi}}+\frac{1}{3H}V'(\hat{\phi}).
\end{equation}\\

\noindent Applying the field split of Eq. (\ref{field_split}), we obtain\\

\begin{equation}
    \label{overdamped_langevin}
    0=\Dot{\overline{\phi}}+\frac{1}{3H}V'(\overline{\phi})-\hat{\xi}(t,\mathbf{x}),
\end{equation}\\
where\\

\begin{equation}
    \label{od_noise}
    \hat{\xi}(t,\mathbf{x})=\epsilon a(t) H^2\int\frac{d^3\mathbf{k}}{(2\pi)^3}\delta(k-\epsilon a(t)H)\hat{\phi}_k(t,\mathbf{x}).
\end{equation}\\

\noindent For massless scalars, the vacuum Bunch-Davies mode solutions \cite{bunch_davies_1978,tagirov_QFT_deSitter,Birrell-Davies_book} are given by\\

\begin{equation}
    \label{massless_field_modes}
    \phi_k(t,\mathbf{x})=-\frac{H}{\sqrt{2k}}\qty(\frac{1}{Ha(t)}+\frac{i}{k})e^{i\qty(\frac{k}{Ha(t)}-\mathbf{k}\cdot\mathbf{x})}.
\end{equation}\\

\noindent The key observation of Starobinsky and Yokoyama~\cite{starobinsky_1986, Starobinsky-Yokoyama_1994} was that Eq. (\ref{overdamped_langevin}) may be approximated by a Langevin equation\\

\begin{equation}
    \label{langevin_eq_OD}
    0=\Dot{\phi}+\frac{1}{3H}V'(\phi)-\xi(t,\mathbf{x}),
\end{equation}\\

\noindent with a stochastic white noise contribution $\xi(t,\mathbf{x})$ that satisfies\\
\begin{equation}
    \label{od_noise_variance}
     \expval{{\xi}(t,\mathbf{x}){\xi}(t',\mathbf{x})}=\lim_{\epsilon\rightarrow 0}
    \expval{\hat{\xi}(t,\mathbf{x})\hat{\xi}(t',\mathbf{x})}=\frac{H^3}{4\pi^2}\delta(t-t').
\end{equation}\\

\noindent Thus, $\phi$ is now considered as a stochastic quantity with an associated probability distribution function and hence correlators can be evaluated by statistical methods.\\

\subsection{The cut-off method}

We will now consider massive fields, requiring us to go beyond the overdamped approximation. The naive extension of Starobinsky and Yokoyama's cut-off approach is to follow the same procedure as in the overdamped limit, but now we also need to split the time derivative of the field \cite{Grain_2017, Pinol:2020}. This gives\\

\begin{equation}
    \label{field_derivative_split}
    \begin{split}
    \hat{\pi}:=\Dot{\hat{\phi}}&=\Dot{\overline{\phi}}+\partial_t\qty(\hat{\delta\phi})\\
    &=\overline{\pi}+\hat{\delta\pi},
    \end{split}
\end{equation}\\

\noindent where $\overline{\pi}$ contains the classical modes and\\

\begin{equation}
    \label{subhorizon_field_derivative}
    \hat{\delta\pi}=\int\frac{d^3\mathbf{k}}{(2\pi)^3}W_k(t)\hat{\pi}_k(t,\mathbf{x}),
\end{equation}\\

\noindent where $\hat{\pi}_k(t,\mathbf{x})=\pi_k(t,\mathbf{x})\hat{a}_k+\pi^*_k(t,\mathbf{x})\hat{a}_k^{\dagger}$ and $W_k(t)$ is a window function that satisfies $W_k(t)=1$ for $k\gg a(t)H$ and $W_k(t)=0$ for $k\ll a(t)H$. The mode functions obey the equation of motion (\ref{2PDE_eom}), which is solved to give\\

\begin{subequations}
\label{mode_functions}
    \begin{equation}
        \label{field_mode}
        \phi_k(t,\mathbf{x})=\sqrt{\frac{\pi}{4H}}a(t)^{-3/2}\mathcal{H}_{\nu}^{(1)}\qty(\frac{k}{a(t)H})e^{-i\mathbf{k}\cdot\mathbf{x}},
    \end{equation}\\
    \begin{small}
    \begin{equation}
        \label{field_derivative_mode}
        \pi_k(t,\mathbf{x})=-\sqrt{\frac{\pi}{16H}}a(t)^{-3/2}\qty[3H\mathcal{H}_{\nu}^{(1)}\qty(\frac{k}{a(t)H})+\frac{k}{a(t)}\qty(\mathcal{H}_{\nu-1}^{(1)}\qty(\frac{k}{a(t)H})-\mathcal{H}_{\nu+1}^{(1)}\qty(\frac{k}{a(t)H}))]e^{-i\mathbf{k}\cdot\mathbf{x}}
    \end{equation}\\
    \end{small}
\end{subequations}

\noindent for the Bunch-Davies vacuum \cite{bunch_davies_1978,tagirov_QFT_deSitter,Birrell-Davies_book}, where $\mathcal{H}_{\nu}^{(i)}(z)$ is the Hankel function of the $i$-th kind and $\nu=\sqrt{\frac{9}{4}-\frac{m^2}{H^2}}$. We can now write the equation of motion in terms of the splits as\\

\begin{small}
\begin{subequations}
    \label{split_eom}
    \begin{equation}
        \label{line1}
        \begin{split}
        \Dot{\overline{\phi}}=\overline{\pi}-\int \frac{d^3\mathbf{k}}{(2\pi)^3}\Dot{W}_k(t)\hat{\phi}_k(t,\mathbf{x}),\\
        \end{split}
    \end{equation}\\
    \begin{equation}
        \label{line2}
        \begin{split}
        \Dot{\overline{\pi}}=-3H\overline{\phi}-V'(\overline{\phi})-\int\frac{d^3\mathbf{k}}{(2\pi)^3}\Dot{W}_k(t)\hat{\pi}_k(t,\mathbf{x}),
        \end{split}
    \end{equation}\\
\end{subequations}
\end{small}
 
\noindent where Eq. (\ref{line1}) and (\ref{line2}) are the first and second rows of Eq. (\ref{2PDE_eom}) respectively. Thus, the equations of motion can be written as a system of first-order differential equations\\

\begin{equation}
    \label{Langevin_equation0}
    \begin{pmatrix}\Dot{\overline{\phi}}\\\Dot{\overline{\pi}}\end{pmatrix}=\begin{pmatrix}\overline{\pi}\\-3H\overline{\pi}-V'(\overline{\phi})\end{pmatrix}+\begin{pmatrix}\hat{\xi}_{\phi}(t,\mathbf{x})\\\hat{\xi}_{\pi}(t,\mathbf{x})\end{pmatrix},
\end{equation}\\

\noindent where\\

\begin{subequations}
    \label{quantum_noise}
    \begin{equation}
        \label{phi_noise}
        \hat{\xi}_{\phi}(t,\mathbf{x})=-\int\frac{d^3\mathbf{k}}{(2\pi)^3}\Dot{W}_k(t)\hat{\phi}_k(t,\mathbf{x}),
    \end{equation}
    \begin{equation}
        \label{pi_noise}
        \hat{\xi}_{\pi}(t,\mathbf{x})=-\int\frac{d^3\mathbf{k}}{(2\pi)^3}\Dot{W}_k(t)\hat{\pi}_k(t,\mathbf{x}).
    \end{equation}\\
\end{subequations}

\noindent Once again, we may approximate Eq. (\ref{Langevin_equation0}) as a Langevin equation, where the field and field momentum move from quantum to classical objects. This is written as\\

\begin{equation}
    \label{Langevin_equation}
    \begin{pmatrix}\Dot{{\phi}}\\\Dot{{\pi}}\end{pmatrix}=\begin{pmatrix}{\pi}\\-3H{\pi}-V'({\phi})\end{pmatrix}+\begin{pmatrix}{\xi}_{\phi}(t,\mathbf{x})\\{\xi}_{\pi}(t,\mathbf{x})\end{pmatrix},
\end{equation}\\

\noindent where the stochastic noise $\xi_i$ satisfies\\

\begin{equation}
     \expval{{\xi}_i(t,\mathbf{x}){\xi}_j(t',\mathbf{x})}= \expval{\hat{\xi}_i(t,\mathbf{x})\hat{\xi}_j(t',\mathbf{x})},
\end{equation}\\

\noindent for  $\{i,j\}=\{\phi,\pi\}$. Note that we will refer to $\pi$ as the field momentum, though this is now slightly more complicated than just the time derivative of the field, as per Eq. (\ref{Langevin_equation0}).\\

Computing the quantum correlator for a sharp cut-off at $k=\epsilon a(t)H$, namely $W_k(t)=\theta(k-\epsilon a(t)H)$, where $\epsilon$ is a parameter whose relevance we will discuss later, we obtain\\

\begin{equation}
    \label{noise_amplitude_langevin}
    \expval{\xi_i(t,\mathbf{x})\xi_j(t',\mathbf{x})}=\sigma^2_{cut,ij}\delta(t-t'),
\end{equation}\\

\noindent where\\

\begin{subequations}
    \label{noise_amplitudes_modes}
    \begin{align}
         \sigma_{cut,\phi\phi}^2=&\frac{\epsilon^3 a(t)^3 H^4}{2\pi^2}\abs{\phi_{\epsilon a(t) H}(t,\mathbf{x})}^2\\
         =&\frac{H^3 \epsilon^3}{8\pi}\mathcal{H}_{\nu}^{(1)}(\epsilon)\mathcal{H}_{\nu}^{(2)}(\epsilon),\nonumber\\&\nonumber\\
        \sigma_{cut,\phi\pi}^2=&\sigma_{cut,\pi\phi}^2=\frac{\epsilon^3 a(t)^3 H^4}{4\pi^2}\qty(\phi_{\epsilon a(t) H}(t,\mathbf{x})\pi^*_{\epsilon a(t)H}(t,\mathbf{x})+\pi_{\epsilon a(t) H}(t,\mathbf{x})\phi^*_{\epsilon a(t)H}(t,\mathbf{x}))\\
        =&-\frac{H^4\epsilon^3}{32\pi}\Bigg[\epsilon\qty(\mathcal{H}_{\nu-1}^{(1)}(\epsilon)-\mathcal{H}_{\nu+1}^{(1)}(\epsilon))\mathcal{H}_{\nu}^{(2)}(\epsilon)\nonumber\\
        &+\mathcal{H}_{\nu}^{(1)}\qty(\epsilon\mathcal{H}_{\nu-1}^{(2)}(\epsilon)+6\mathcal{H}_{\nu}^{(2)}(\epsilon)-\epsilon\mathcal{H}_{\nu+1}^{(2)}(\epsilon))\Bigg],\nonumber\\&\nonumber\\
        \sigma_{cut,\pi\pi}^2&=\frac{\epsilon^3 a(t)^3 H^4}{2\pi^2}\abs{\pi_{\epsilon a(t) H}(t,\mathbf{x})}^2\\
        =&\frac{H^5\epsilon^3}{32\pi}\qty(\epsilon\mathcal{H}_{\nu-1}^{(1)}(\epsilon)+3\mathcal{H}_{\nu}^{(1)}(\epsilon)-\epsilon\mathcal{H}_{\nu+1}^{(1)}(\epsilon))\qty(\epsilon\mathcal{H}_{\nu-1}^{(2)}(\epsilon)+3\mathcal{H}_{\nu}^{(2)}(\epsilon)-\epsilon\mathcal{H}_{\nu+1}^{(2)}(\epsilon)),\nonumber
    \end{align}\\
\end{subequations}

\noindent and the subscript $cut$ stands for cut-off. We note that $\sigma_{cut,\phi\pi}^2$ is given by the classical part of the quantum noise correlator as we know that it must be real for this to be considered a stochastic process. Choosing $\epsilon\ll 1$ ensures that the classical symmetric part dominates over the quantum antisymmetric part. The details of this statement are dealt with more carefully in Ref.~\cite{Grain_2017,Pinol:2020}.\\

\subsection{The Fokker-Planck equation}
\label{subsec:fokker-planck}

The time evolution of the probability distribution function (PDF) $P(\phi,\pi;t)$ of the variables $\phi$ and $\pi$ is given by the Fokker-Planck equation associated with the Langevin equation (\ref{Langevin_equation})\\

\begin{equation}
    \label{fokker-planck_phi-pi}
    \begin{split}
    \pdv{P(\phi,\pi;t)}{t}=&3HP(\phi,\pi;t)-\pi\pdv{P(\phi,\pi;t)}{\phi}+\qty(3H \pi+V'(\phi))\pdv{P(\phi,\pi;t)}{\pi}\\&+\frac{1}{2}\sigma_{\phi\phi}^2\pdv[2]{P(\phi,\pi;t)}{\phi}+\sigma_{\phi\pi}^2\frac{\partial^2P(\phi,\pi;t)}{\partial\phi\partial\pi}+\frac{1}{2}\sigma_{\pi\pi}^2\pdv[2]{P(\phi,\pi;t)}{\pi},
    \end{split}
\end{equation}\\

\noindent where, in this Section, we consider general white noise with the form\\

\begin{equation}
    \label{noise_amplitude_langevin_gen}
    \expval{\xi_i(t,\mathbf{x})\xi_j(t',\mathbf{x})}=\sigma_{ij}\delta(t-t'),
\end{equation}\\

\noindent which does not necessarily have the amplitude (\ref{noise_amplitudes_modes}).\\

In principle it is possible to solve Eq.~(\ref{fokker-planck_phi-pi}) using the spectral expansion in the same way as in the overdamped limit ~\cite{Markkanen_2019}.
However, the two-dimensional eigenvalue equation is hard to solve numerically. It is also not self-adjoint, and therefore the eigenfunctions are not orthogonal. In this paper, we will restrict ourselves to the free field case,\\

\begin{equation}
\label{equ:freepot}
    V(\phi)=\frac{1}{2}m^2\phi^2,
\end{equation}\\

\noindent where we can solve Eq.~(\ref{fokker-planck_phi-pi}) using a different technique. It is worth noting, that in the special case $\sigma_{\phi\phi}^2=\sigma_{\phi\pi}^2=0$, the eigenvalue equation can also be solved analytically using creation and annihilation operators~\cite{pavliotis_stochastic_apps}.\\

For the free potential (\ref{equ:freepot}), the equilibrium solution is\\

\begin{equation}
    \label{1PDF_equilibrium_solution}
    P_{eq}(\phi,\pi)\propto 
    e^{-\frac{3H\qty(((9H^2+m^2)\sigma_{\phi\phi}^2+6H\sigma_{\phi\pi}^2+\sigma_{\pi\pi}^2)\pi^2+6Hm^2\sigma_{\phi\phi}^2\phi\pi+(m^2\sigma_{\phi\phi}^2+\sigma_{\pi\pi}^2)m^2\phi^2)}{(m^2\sigma_{\phi\phi}^2+3H\sigma_{\phi\pi}^2+\sigma_{\pi\pi}^2)^2+9H^2(\sigma_{\phi\phi}^2\sigma_{\pi\pi}^2-\sigma_{\phi\pi}^4)}},
\end{equation}\\

\noindent with the normalisation condition $\int d\phi\int d\pi P_{eq}(\phi,\pi)=1$.\\

To obtain the time-dependent solution, we introduce new dynamical variables $(q,p)$ as\\

\begin{equation}
    \label{phi,pi-->q,p}
    \begin{pmatrix}p\\q\end{pmatrix}=\frac{1}{\sqrt{1-\frac{\alpha}{\beta}}}\begin{pmatrix}1&\alpha H\\\frac{1}{\beta H}&1\end{pmatrix}\begin{pmatrix}\pi\\\phi\end{pmatrix},
\end{equation}\\

\noindent where $\alpha=\frac{3}{2}-\nu$ and $\beta=\frac{3}{2}+\nu$, with $\nu=\sqrt{\frac{9}{4}-\frac{m^2}{H^2}}$. The inverse transformation is given by\\

\begin{equation}
    \label{p,q-->pi,phi}
    \begin{pmatrix}\pi\\\phi\end{pmatrix}=\frac{1}{\sqrt{1-\frac{\alpha}{\beta}}}\begin{pmatrix}1&-\alpha H\\-\frac{1}{\beta H}&1\end{pmatrix}\begin{pmatrix}p\\q\end{pmatrix}.
\end{equation}\\

\noindent In these new variables, the Langevin equation (\ref{Langevin_equation}) can be written as\\

\begin{equation}
    \label{langevin_p,q}
    \begin{pmatrix}\Dot{q}\\\Dot{p}\end{pmatrix}=\begin{pmatrix}-\alpha H q\\-\beta H p\end{pmatrix}+\begin{pmatrix}\xi_q\\\xi_p\end{pmatrix},
\end{equation}\\

\noindent where $\xi_q=\frac{1}{\sqrt{1-\frac{\alpha}{\beta}}}\qty(\frac{1}{\beta H}\xi_{\pi}+\xi_{\phi})$ and $\xi_p=\frac{1}{\sqrt{1-\frac{\alpha}{\beta}}}\qty(\xi_{\pi}+\alpha H \xi_{\phi})$. Thus, we have two 1-dimensional Langevin equations with correlated noise. The resulting Fokker-Planck equation for the PDF $P(q,p;t)$ is\\

\begin{equation}
    \label{p,q_fokker-planck}
    \begin{split}
    \pdv{P(q,p;t)}{t}=&\alpha H P(q,p;t)+\alpha H q\pdv{P(q,p;t)}{q}+\frac{1}{2}\sigma_{qq}^2\pdv[2]{P(q,p;t)}{q}\\&+\beta H P(q,p;t)+\beta H p\pdv{P(q,p;t)}{p}+\frac{1}{2}\sigma_{pp}^2\pdv[2]{P(q,p;t)}{p}\\&+\sigma_{qp}^2\frac{\partial^2P(q,p;t)}{\partial q\partial p} ,
    \end{split}
\end{equation}\\

\noindent where $\sigma_{qq}^2$, $\sigma_{qp}^2$ and $\sigma_{pp}^2$ are defined in the same way by Eq. (\ref{noise_amplitude_langevin_gen}), but now $(i,j)\in\{q,p\}$. The equilibrium solution to this Fokker-Planck equation is\\

\begin{equation}
    \label{p,q_equilibrium_1PDF}
    P_{eq}(q,p)=\frac{3H}{\pi}\sqrt{\frac{\alpha \beta}{9\sigma_{qq}^2\sigma_{pp}^2-4\alpha\beta \sigma_{qp}^4}}e^{-\frac{9H(\sigma_{qq}^2\beta p^2-\frac{4}{3}\sigma_{qp}^2\alpha\beta q p+\sigma_{pp}^2\alpha q^2)}{9\sigma_{qq}^2\sigma_{pp}^2-4\alpha\beta \sigma_{qp}^4}},
\end{equation}\\

\noindent where we have included the normalisation found by the condition $\int dp \int dq P_{eq}(q,p)=1$.\\

For completeness, we note that the transformation to and from the $(\phi,\pi)$ noise amplitudes with respect to the $(q,p)$ noise amplitudes is respectively\\

\begin{subequations}
    \label{noise_amplitudes_transformation}
    \begin{equation}
        \label{sigma_q,p-->sigma_phi,pi}
        \begin{split}
            \sigma_{qq}^2=&\frac{1}{1-\frac{\alpha}{\beta}}\qty(\frac{1}{\beta^2H^2}\sigma_{\pi\pi}^2+\frac{2}{\beta H}\sigma_{\phi\pi}^2+\sigma_{\phi\phi}^2),\\
            \sigma_{qp}^2=&\frac{1}{1-\frac{\alpha}{\beta}}\qty(\frac{1}{\beta H}\sigma_{\pi\pi}^2+\qty(1+\frac{\alpha}{\beta })\sigma_{\phi\pi}^2+\alpha H\sigma_{\phi\phi}^2),\\
            \sigma_{pp}^2=&\frac{1}{1-\frac{\alpha}{\beta}}\qty(\sigma_{\pi\pi}^2+2\alpha H\sigma_{\phi\pi}^2+\alpha^2H^2\sigma_{\phi\phi}^2);
        \end{split}
    \end{equation}\\
    \begin{equation}
        \label{sigma_phi,pi-->sigma_q,p}
        \begin{split}
            \sigma_{\phi\phi}^2=&\frac{1}{1-\frac{\alpha}{\beta}}\qty(\frac{1}{\beta^2H^2}\sigma_{pp}^2-\frac{2}{\beta H}\sigma_{qp}^2+\sigma_{qq}^2),\\
            \sigma_{\phi\pi}^2=&\frac{1}{1-\frac{\alpha}{\beta}}\qty(-\frac{1}{\beta H}\sigma_{pp}^2+\qty(1+\frac{\alpha}{\beta })\sigma_{qp}^2-\alpha H\sigma_{qq}^2),\\
            \sigma_{\pi\pi}^2=&\frac{1}{1-\frac{\alpha}{\beta}}\qty(\sigma_{pp}^2-2\alpha H\sigma_{qp}^2+\alpha^2H^2\sigma_{qq}^2).   
        \end{split}
    \end{equation}\\
\end{subequations}

\section{Evaluation of correlators}
\label{sec:qft_comparison}

\subsection{Stochastic correlation functions}
\label{subsec:correlators}

We will now evaluate the $(\phi,\pi)$ correlators in terms of the $(q,p)$ correlators using Eq. (\ref{p,q-->pi,phi}). For the stochastic approach, we evaluate the correlators at equal points in space before transforming to equal-time correlators at the end of this Section. The equal-space $(\phi,\pi)$ correlators are written as\\

\begin{subequations}
\label{phi,pi_correlators_pq}
\begin{align}
        \label{phi-phi_pq}
        \expval{\phi(0)\phi(t)}&=\frac{1}{1-\frac{\alpha}{\beta}}\qty(\frac{1}{\beta^2H^2}\expval{p(0)p(t)}-\frac{1}{\beta H}\qty(\expval{q(0)p(t)}+\expval{p(0)q(t)})+\expval{q(0)q(t)}),\\
        \label{phi-pi_pq}
        \expval{\phi(0)\pi(t)}&=\frac{1}{1-\frac{\alpha}{\beta}}\qty(-\frac{1}{\beta H}\expval{p(0)p(t)}+\expval{q(0)p(t)}+\frac{\alpha}{\beta}\expval{p(0)q(t)}-\alpha H\expval{q(0)q(t)}),\\
        \label{pi-phi_pq}
        \expval{\pi(0)\phi(t)}&=\frac{1}{1-\frac{\alpha}{\beta}}\qty(-\frac{1}{\beta H}\expval{p(0)p(t)}+\frac{\alpha}{\beta}\expval{q(0)p(t)}+\expval{p(0)q(t)}-\alpha H\expval{q(0)q(t)}),\\
        \label{pi-pi_pq}
        \expval{\pi(0)\pi(t)}&=\frac{1}{1-\frac{\alpha}{\beta}}\qty(\expval{p(0)p(t)}-\alpha H\qty(\expval{q(0)p(t)}+\expval{p(0)q(t)})+\alpha^2H^2\expval{q(0)q(t)}).
\end{align}\\
\end{subequations}

\noindent To calculate the $(q,p)$ correlators, we introduce a time-evolution operator $U(q_0,q,p_0,p;t)$, which is defined as the Green's function of the Fokker-Planck equation and therefore obeys\\

\begin{equation}
    \label{fokker-planck_time-ev_op}
    \begin{split}
    \pdv{U(q_0,q,p_0,p;t)}{t}=&\alpha H U(q_0,q,p_0,p;t)+\alpha H q\pdv{U(q_0,q,p_0,p;t)}{q}+\frac{1}{2}\sigma_{qq}^2\pdv[2]{U(q_0,q,p_0,p;t)}{q}\\&+\beta H U(q_0,q,p_0,p;t)+\beta H p\pdv{U(q_0,q,p_0,p;t)}{p}+\frac{1}{2}\sigma_{pp}^2\pdv[2]{U(q_0,q,p_0,p;t)}{p}\\&+\sigma_{qp}^2\frac{\partial^2U(q_0,q,p_0,p;t)}{\partial q\partial p} 
    \end{split},    
\end{equation}

\noindent for all values of $q_0$ and $p_0$. Then, the time-dependence of the PDF is given by\\

\begin{equation}
    \label{time-ev_definition}
    P(q,p;t)=\int dp_0\int dq_0 P(q_0,p_0;0)U(q_0,q,p_0,p;t),
\end{equation}\\

\noindent such that the time-evolution operator obeys the Fokker-Planck equation (\ref{p,q_fokker-planck}). Unfortunately, this is not an easy equation to solve. However, if we initially only consider the 2-point correlation functions such as those in Eq. (\ref{phi,pi_correlators_pq}), we only need to evolve $p$ or $q$ forward in time for any given correlator, not both simultaneously. Therefore, one only needs to use the 1-dimensional time evolution operators $U_q(q_0,q;t)$ and $U_p(p_0,p;t)$
defined by\\

\begin{subequations}
    \label{time-ev_1PDF}
    \begin{equation}
        \label{q_time-ev_1pdf}
        P_q(q;t)=\int dq_0 P_q(q_0;0)U_q(q_0,q;t),
    \end{equation}
    \begin{equation}
        \label{p_time-ev_1pdf}
        P_p(p;t)=\int dp_0 P_p(p_0;0)U_p(p_0,p;t),
    \end{equation}\\
\end{subequations}

\noindent respectively, where $P_q(q;t)$ and $P_p(p;t)$ are the time-dependent univariate probability distributions.
These time-evolution operators satisfy the Fokker-Planck equations\\

\begin{subequations}
    \label{time-ev_pde}
    \begin{align}
        \label{q_time-ev_pde}
        \pdv{U_q(q_0,q;t)}{t}&=\alpha H U_q(q_0,q;t)+\alpha H q \pdv{U_q(q_0,q;t)}{q}+\frac{1}{2}\sigma_{qq}^2\pdv[2]{U_q(q_0,q;t)}{q},\\[10pt]
        \label{p_time-ev_pde}
        \pdv{U_p(p_0,p;t)}{t}&=\beta H U_p(p_0,p;t)+\beta H p \pdv{U_p(p_0,p;t)}{p}+\frac{1}{2}\sigma_{pp}^2\pdv[2]{U_p(p_0,p;t)}{p},
    \end{align}\\
\end{subequations}
which can be derived in the standard way from the two components of the Langevin equation~(\ref{langevin_p,q}).\\

Alternatively, they can also be obtained from the two-dimensional Fokker-Planck equation (\ref{fokker-planck_time-ev_op}), by first observing that they can be expressed in terms of the two-dimensional time-evolution operators as\\

\begin{subequations}
    \begin{align}
        U_q(q_0,q;t)&=\int dp\, U(q_0,q,p_0,p;t),\\
        U_p(p_0,p;t)&=\int dq\, U(q_0,q,p_0,p;t),
    \end{align}\\
\end{subequations}

which do not depend on $p_0$ and $q_0$, respectively.
Integrating Eq.~(\ref{fokker-planck_time-ev_op}) over $p$ and $q$, respectively, and integrating the relevant terms by parts gives Eqs.~(\ref{q_time-ev_pde}) and (\ref{p_time-ev_pde}).\\

It is a special property of the coordinates $(q,p)$ that one obtains time-evolution equations that only depend on one variable. It comes as a result of the Langevin equation (\ref{langevin_p,q}) being diagonal i.e. the time derivative of $q$ is not dependent on $p$ and, similarly, the time derivative of $p$ is not dependent on $q$. It is this property that allows us to analytically evaluate the 2-point correlators using the 1-dimensional time-evolution operators. Note, however, that in order to calculate the higher-order correlators, one needs to evaluate the full time-evolution operator.\\

Eq. (\ref{time-ev_pde}) can be solved via a spectral expansion \cite{Markkanen_2019} as outlined in Appendix \ref{app:eigenspectrum}. The resulting solutions are\\

\begin{subequations}
    \label{time-ev_solution}
    \begin{equation}
        \label{q_time-ev_solution}
        U_q(q_0,q;t)=e^{-\frac{\alpha H}{2\sigma_{qq}^2}(q^2-q_0^2)}\sum_n Q_n(q)Q_n(q_0)e^{-nH\alpha t},
    \end{equation}
    \begin{equation}
        \label{p_time-ev_solution}
        U_p(p_0,p;t)=e^{-\frac{\beta H}{2\sigma_{pp}^2}(p^2-p_0^2)}\sum_n P_n(p)P_n(p_0)e^{-nH\beta t},
    \end{equation}\\
\end{subequations}
\noindent where\\

\begin{subequations}
\label{eigenfunctions}
    \begin{equation}
        \label{Qn}
        Q_n(q)=\frac{1}{\sqrt{2^n n!}}\qty(\frac{\alpha H}{\pi \sigma_{qq}^2})^{\frac{1}{4}}H_n\qty(\sqrt{\frac{\alpha H}{\sigma_{qq}^2}}q)e^{-\frac{\alpha H}{2\sigma_{qq}^2}q^2},
    \end{equation}
    \begin{equation}
        \label{Pn}
        P_n(p)=\frac{1}{\sqrt{2^n n!}}\qty(\frac{\beta H}{\pi \sigma_{pp}^2})^{\frac{1}{4}}H_n\qty(\sqrt{\frac{\beta H}{\sigma_{pp}^2}}p)e^{-\frac{\beta H}{2\sigma_{pp}^2}p^2},
    \end{equation}\\
\end{subequations}

\noindent where $H_n(z)$ is the Hermite polynomial.\\

\subsubsection{2-point equal-space stochastic correlators}

The 2-point $(q,p)$ correlators are given by\\ 

\begin{subequations}
\label{p,q-correlators_time-ev}
\begin{align}
        \expval{q(0)q(t)}&=\int dp_0\int dq\int dq_0 P_{eq}(q_0,p_0)U_q(q_0,q;t)q_0 q=\frac{\sigma_{qq}^2}{2\alpha H}e^{-\alpha Ht},\\
        \expval{p(0)q(t)}&=\int dp_0\int dq\int dq_0 P_{eq}(q_0,p_0)U_q(q_0,q;t)p_0 q=\frac{\sigma_{qp}^2}{3H}e^{-\alpha Ht},\\
        \expval{q(0)p(t)}&=\int dq_0 \int dp \int dp_0 P_{eq}(q_0,p_0)U_p(p_0,p;t)q_0 p=\frac{\sigma_{qp}^2}{3H}e^{-\beta Ht},\\
        \expval{p(0)p(t)}&=\int dp\int dp_0\int dq_0 P_{eq}(q_0,p_0)U_p(p_0,p;t)p_0p=\frac{\sigma_{pp}^2}{2\beta H}e^{-\beta Ht}.
\end{align}\\
\end{subequations}

\noindent and the equal-space $(\phi,\pi)$ stochastic correlators are calculated using Eq. (\ref{phi,pi_correlators_pq}) as\\

\begin{subequations}
\label{equal-space_(phi,pi)_stochastic_correlators}
\begin{align}
    \label{timelike_phi-phi_stochastic_correlator}
    \expval{\phi(0)\phi(t)}&=\frac{1}{1-\frac{\alpha}{\beta}}\qty[\qty(\frac{\sigma_{qq}^2}{2H\alpha}-\frac{\sigma_{qp}^2}{3H^2\beta})e^{-\alpha H t}+\qty(\frac{\sigma_{pp}^2}{2H^3\beta^3}-\frac{\sigma_{qp}^2}{3H^2\beta})e^{-\beta H t}],\\
    \label{timelike_phi-pi_stochastic_correlator}
    \expval{\phi(0)\pi(t)}&=\frac{1}{1-\frac{\alpha}{\beta}}\qty[\qty(-\frac{\sigma_{qq}^2}{2}+\frac{\alpha\sigma_{qp}^2}{3H\beta})e^{-\alpha H t}+\qty(-\frac{\sigma_{pp}^2}{2H^2\beta^2}+\frac{\sigma_{qp}^2}{3H})e^{-\beta H t}],\\
    \label{timelike_pi-phi_stochastic_correlator}
    \expval{\pi(0)\phi(t)}&=\frac{1}{1-\frac{\alpha}{\beta}}\qty[\qty(-\frac{\sigma_{qq}^2}{2}+\frac{\sigma_{qp}^2}{3H})e^{-\alpha H t}+\qty(-\frac{\sigma_{pp}^2}{2H^2\beta^2}+\frac{\alpha\sigma_{qp}^2}{3H\beta})e^{-\beta H t}],\\
    \label{timelike_pi-pi_stochastic_correlator}
    \expval{\pi(0)\pi(t)}&=\frac{1}{1-\frac{\alpha}{\beta}}\qty[\qty(\frac{\alpha H\sigma_{qq}^2}{2}-\frac{\alpha\sigma_{qp}^2}{3})e^{-\alpha H t}+\qty(\frac{\sigma_{pp}^2}{2H\beta}-\frac{\alpha\sigma_{qp}^2}{3})e^{-\beta H t}].
\end{align}\\
\end{subequations}

\subsubsection{Stochastic variance}
\label{subsubsec:stoch_variance}

\noindent For completeness, we can also evaluate the stochastic variances. We will discuss these in the context of the quantum result in Section \ref{subsec:variance}. For now, they are given by\\

\begin{subequations}
\label{stochastic_variance}
    \begin{align}
        \label{stoch_phiphi_variance}
        \expval{\phi^2}&=\frac{1}{1-\frac{\alpha}{\beta}}\qty(\expval{q^2}-\frac{2}{\beta H}\expval{qp}+\frac{1}{\beta^2H^2}\expval{p^2}),\\
        \label{stoch_phipi_variance}
        \expval{\phi\pi}&=\frac{1}{1-\frac{\alpha}{\beta}}\qty(-\frac{1}{\beta H}\expval{p^2}+\qty(1+\frac{\alpha}{\beta})\expval{qp}-\alpha H \expval{q^2}),\\
        \label{stoch_pipi_variance}
        \expval{\pi^2}&=\frac{1}{1-\frac{\alpha}{\beta}}\qty(\expval{p^2}-2\alpha H \expval{qp}+\alpha^2H^2\expval{q^2}),
    \end{align}\\
\end{subequations}

\noindent where

\begin{subequations}
    \label{p,q_variance}
    \begin{align}
        \label{<q^2>}
        \expval{q^2}&=\int dq\int dp P_{eq}(p,q)q^2=\frac{\sigma_{qq}^2}{2H\alpha},\\
        \label{<pq>}
        \expval{qp}&=\int dq \int dp P_{eq}(q,p)pq=\frac{\sigma_{qp}^2}{3H},\\
        \label{<p^2}
        \expval{p^2}&=\int dq\int dp P_{eq}(q,p)p^2=\frac{\sigma_{pp}^2}{2H\beta}.
    \end{align}\\
\end{subequations}

\noindent Thus, the stochastic variances in terms of a general noise term are given by\\

\begin{subequations}
    \label{stoch_variance_noise}
    \begin{align}
        \label{stoch_phiphi_variance_noise}
        \expval{\phi^2}&=\frac{1}{1-\frac{\alpha}{\beta}}\qty(\frac{\sigma_{qq}^2}{2H \alpha}-\frac{2\sigma_{qp}^2}{3\beta H^2}+\frac{\sigma_{pp}^2}{2\beta^3H^3}),\\
        \label{stoch_phipi_variance_noise}
        \expval{\phi\pi}&=\frac{1}{1-\frac{\alpha}{\beta}}\qty(-\frac{\sigma_{qq}^2}{2}+\qty(1+\frac{\alpha}{\beta})\frac{\sigma_{qp}^2}{3H}-\frac{\sigma_{pp}^2}{2\beta^2H^2}),\\
        \label{stoch_pipi_variance_noise}
        \expval{\pi^2}&=\frac{1}{1-\frac{\alpha}{\beta}}\qty(\frac{\alpha H \sigma_{qq}^2}{2}-\frac{2\alpha\sigma_{qp}^2}{3}+\frac{\sigma_{pp}^2}{2\beta H}).
    \end{align}\\
\end{subequations}

\subsubsection{Joint unequal-time probability distribution function}

Now that we have calculated all the timelike 2-point $(\phi,\pi)$ stochastic correlators and their associated variance, we can write the joint probability that $\phi(0)=\phi_0$, $\pi(0)=\pi_0$, $\phi(t)=\phi$ and $\pi(t)=\pi$, for any given $\phi_0,\pi_0,\phi,\pi$. This is given by the joint unequal-time PDF as\\

\begin{equation}
    \label{joint_PDF_time}
    P_2(\phi_0,\phi,\pi_0,\pi;t)= N(t) e^{-\frac{1}{2}\Phi^T M(t)^{-1}\Phi}
\end{equation}

\noindent where

\begin{subequations}
    \label{matrix&vector}
    \begin{equation}
        \label{vector}
        \Phi=(\pi_0,\phi_0,\pi,\phi)^T,
    \end{equation}\\
    \begin{equation}
        \label{matrix}
        M(t)=\begin{pmatrix}\expval{\pi_0\pi_0}&\expval{\pi_0\phi_0}&\expval{\pi_0\pi}&\expval{\pi_0\phi}\\\expval{\phi_0\pi_0}&\expval{\phi_0\phi_0}&\expval{\phi_0\pi}&\expval{\phi_0\phi}\\\expval{\pi\pi_0}&\expval{\pi\phi_0}&\expval{\pi\pi}&\expval{\pi\phi}\\\expval{\phi\pi_0}&\expval{\phi\phi_0}&\expval{\phi\pi}&\expval{\phi\phi}\end{pmatrix}
    \end{equation}\\
\end{subequations}

\noindent The normalisation constant $N(t)$ is found by the condition $\int d\phi \int d\phi_0 \int d\pi \int d\pi_0 P_2(\phi_0,\phi,\pi_0,\pi;t)=1$. Note that Eq. (\ref{joint_PDF_time}) is a solution to the Fokker-Planck equation (\ref{fokker-planck_phi-pi}) for any value of $\phi_0$ and $\pi_0$. This allows us to calculate higher-order correlators directly, as will be shown in Section \ref{subsec:higher-order_correlators}.\\

One can also write the joint unequal-time PDF in terms of the time-evolution operator as\\

\begin{equation}
    \label{joint_PDF_2-point}
    P_2(\phi_0,\phi,\pi_0,\pi;t)=P_{eq}(\phi_0,\pi_0)U\qty(q(\phi_0,\pi_0),q(\phi,\pi),p(\phi_0,\pi_0),p(\phi,\pi);t)
\end{equation}\\

\noindent where $q$ and $p$ are written in terms of $\phi$ and $\pi$ as of Eq. (\ref{phi,pi-->q,p}).\\

\subsubsection{2-point equal-time stochastic correlators}

The quantum objects that are of interest physically are the equal-time correlators. Currently, the stochastic correlators are for equal points in space and thus we need to find a way of moving to equal times. Following the original work by Starobinsky and Yokoyama \cite{Starobinsky-Yokoyama_1994}, we introduce a time coordinate\\

\begin{equation}
    \label{spatial_time_coord}
    t_r=-\frac{1}{H}\ln\qty(Ha(t)\abs{\mathbf{x}-\mathbf{x}'}).
\end{equation}\\

\noindent We then connect the equal-space and equal-time correlators by evaluating a 3-point function, incorporating this time coordinate. We will follow a similar line of reasoning as the equal-space correlators, whereby we calculate the equal-time $(q,p)$ correlators and then move to the equal-time $(\phi,\pi)$ correlators using Eq. (\ref{phi,pi_correlators_pq}). The equal-time $(q,p)$ correlators are given by\\

\begin{subequations}
    \begin{align}
        \begin{split}
            \expval{q(t,\mathbf{x})q(t,\mathbf{x}')}&=\int dq_r \int dp_r P_{eq}(q_r,p_r)\int dq U_q(q_r,q;t-t_r)q\int dq' U_q(q_r,q';t-t_r)q'\\
            &=\frac{\sigma_{qq}^2}{2\alpha H}\qty(Ha(t)\abs{\mathbf{x}-\mathbf{x}'})^{-2\alpha}
        \end{split}\\
        \begin{split}
            \expval{p(t,\mathbf{x})q(t,\mathbf{x}')}&=\int dq_r \int dp_r P_{eq}(q_r,p_r)\int dp U_p(p_r,p;t-t_r)p\int dq' U_q(q_r,q';t-t_r)q'\\
            &=\frac{\sigma_{qp}^2}{3H}\qty(Ha(t)\abs{\mathbf{x}-\mathbf{x}'})^{-3}
        \end{split}\\
        \begin{split}
            \expval{q(t,\mathbf{x})p(t,\mathbf{x}')}&=\int dq_r \int dp_r P_{eq}(q_r,p_r)\int dq U_q(q_r,q;t-t_r)q\int dp' U_p(p_r,p';t-t_r)p'\\
            &=\frac{\sigma_{qp}^2}{3 H}\qty(Ha(t)\abs{\mathbf{x}-\mathbf{x}'})^{-3}
        \end{split}\\
        \begin{split}
            \expval{p(t,\mathbf{x})p(t,\mathbf{x}')}&=\int dq_r \int dp_r P_{eq}(q_r,p_r)\int dp U_p(p_r,p;t-t_r)p\int dp' U_p(p_r,p';t-t_r)\\
            &=\frac{\sigma_{pp}^2}{2\beta H}\qty(Ha(t)\abs{\mathbf{x}-\mathbf{x}'})^{-2\beta}
        \end{split}
    \end{align}\\
\end{subequations}

\noindent where $U_q(q_r,q;t-t_r)$ and $U_p(p_r,p;t-t_r)$ are given in Eq. (\ref{phi,pi_correlators_pq}), $q_r=q(t_r)$ and $p_r=p(t_r)$. Thus, using the equal-time version of Eq. (\ref{time-ev_definition}), the equal-time $(\phi,\pi)$ stochastic correlators are given by\\

\begin{subequations}
\label{equal-time_(phi,pi)_stochastic_correlators}
\begin{align}
    \label{spacelike_phi-phi_stochastic_correlator}
    \expval{\phi(t,\mathbf{0})\phi(t,\mathbf{x})}&=\frac{1}{1-\frac{\alpha}{\beta}}\qty[\frac{\sigma_{qq}^2}{2H\alpha}\abs{Ha(t)\mathbf{x}}^{-2\alpha}+\frac{\sigma_{pp}^2}{2H^3\beta^3}\abs{Ha(t)\mathbf{x}}^{-2\beta}-\frac{2\sigma_{qp}^2}{3H^2\beta}\abs{Ha(t)\mathbf{x}}^{-3}],\\
    \label{spacelike_phi-pi_stochastic_correlator}
    \expval{\phi(t,\mathbf{0})\pi(t,\mathbf{x})}&=\frac{1}{1-\frac{\alpha}{\beta}}\qty[-\frac{\sigma_{qq}^2}{2}\abs{Ha(t)\mathbf{x}}^{-2\alpha}-\frac{\sigma_{pp}^2}{2H^2\beta^2}\abs{Ha(t)\mathbf{x}}^{-2\beta}+\frac{\sigma_{qp}^2}{H\beta}\abs{Ha(t)\mathbf{x}}^{-3}],\\
    \label{spacelike_pi-phi_stochastic_correlator}
    \expval{\pi(t,\mathbf{0})\phi(t,\mathbf{x})}&=\frac{1}{1-\frac{\alpha}{\beta}}\qty[-\frac{\sigma_{qq}^2}{2}\abs{Ha(t)\mathbf{x}}^{-2\alpha}-\frac{\sigma_{pp}^2}{2H^2\beta^2}\abs{Ha(t)\mathbf{x}}^{-2\beta}+\frac{\sigma_{qp}^2}{H\beta}\abs{Ha(t)\mathbf{x}}^{-3}],\\
    \label{spacelike_pi-pi_stochastic_correlator}
    \expval{\pi(t,\mathbf{0})\pi(t,\mathbf{x})}&=\frac{1}{1-\frac{\alpha}{\beta}}\qty[\frac{\alpha H\sigma_{qq}^2}{2}\abs{Ha(t)\mathbf{x}}^{-2\alpha}+\frac{\sigma_{pp}^2}{2H\beta}\abs{Ha(t)\mathbf{x}}^{-2\beta}-\frac{2\alpha\sigma_{qp}^2}{3}\abs{Ha(t)\mathbf{x}}^{-3}].
\end{align}\\
\end{subequations}

\subsection{Quantum correlators}
\label{subsec:field_correlator}

In order to understand how effective the stochastic approach is, we must compare the correlators with their QFT counterparts. The simplest correlator to calculate is the 2-point Feynman propagator of the field. In free field theory, this can be evaluated exactly by solving the linear mode functions as \cite{Birrell-Davies_book,tagirov_QFT_deSitter, chernikov...de_Sitter_QFT, bunch_davies_1978}\\

\begin{equation}
    \label{qft_field_propagator}
    \begin{split}
    i\Delta(t,t',\mathbf{x},\mathbf{x}'):=&\expval{\hat{T}\hat{\phi}(t,\mathbf{x})\hat{\phi}(t',\mathbf{x}')}\\=&\frac{H^2}{16\pi^2}\Gamma\qty(\frac{3}{2}+\nu)\Gamma\qty(\frac{3}{2}-\nu){_2}F_1\qty(\frac{3}{2}+\nu,\frac{3}{2}-\nu,2;1+\frac{y}{2})
    \end{split}
\end{equation}\\

\noindent in the Bunch-Davies vacuum, where $_2F_1(a,b,c;z)$ is the hypergeometric function, $\Gamma(z)$ are the Euler-Gamma functions and $y$ is given by\\

\begin{equation}
    \label{de_sitter_inv_quantity}
    y=\cosh(H(t-t'))-\frac{H^2}{2}e^{H(t+t')}\abs{\mathbf{x}-\mathbf{x}'}^2-1.
\end{equation}\\

\noindent From Eq. (\ref{qft_field_propagator}), we can find the quantum $\phi-\pi$, $\pi-\phi$ and $\phi-\phi$ correlators by taking time derivatives. The results are\\

\begin{subequations}
    \begin{align}
        \begin{split}
            \expval{\hat{T}\hat{\phi}(t,\mathbf{x})\hat{\pi}(t',\mathbf{x}')}=&\partial_{t'}\qty(i\Delta(t,t',\mathbf{x},\mathbf{x}'))\\
            =&\qty(-H\sinh(H(t-t'))-\frac{H^3}{2}e^{H(t+t')}\abs{\mathbf{x}-\mathbf{x}'}^2)\\&\times\frac{H^2}{64\pi^2}\Gamma\qty(\frac{5}{2}+\nu)\Gamma\qty(\frac{5}{2}-\nu){_2}F_1\qty(\frac{5}{2}-\nu,\frac{5}{2}+\nu,3;1+\frac{y}{2})\\&
        \end{split}\\
        \begin{split}
            \expval{\hat{T}\hat{\pi}(t,\mathbf{x})\hat{\phi}(t',\mathbf{x}')}=&\partial_{t}\qty(i\Delta(t,t',\mathbf{x},\mathbf{x}'))\\
            =&\qty(H\sinh(H(t-t'))-\frac{H^3}{2}e^{H(t+t')}\abs{\mathbf{x}-\mathbf{x}'}^2)\\&\times\frac{H^2}{64\pi^2}\Gamma\qty(\frac{5}{2}+\nu)\Gamma\qty(\frac{5}{2}-\nu){_2}F_1\qty(\frac{5}{2}-\nu,\frac{5}{2}+\nu,3;1+\frac{y}{2}),\\&
        \end{split}\\
        \begin{split}
            \expval{\hat{T}\hat{\pi}(t,\mathbf{x})\hat{\pi}(t',\mathbf{x}')}=&\partial_t\partial_{t'}\qty(i\Delta(t,t',\mathbf{x},\mathbf{x}'))\\
            =&\qty(-H^2\cosh(H(t-t'))-\frac{H^4}{2}e^{H(t+t')}\abs{\mathbf{x}-\mathbf{x}'}^2)\\&\times\frac{H^2}{64\pi^2}\Gamma\qty(\frac{5}{2}+\nu)\Gamma\qty(\frac{5}{2}-\nu){_2}F_1\qty(\frac{5}{2}-\nu,\frac{5}{2}+\nu,3;1+\frac{y}{2}),
            \\&+\qty(H\sinh(H(t-t'))-\frac{H^3}{2}e^{H(t+t')}\abs{\mathbf{x}-\mathbf{x}'}^2)\\&\times\frac{H^2}{384\pi^2}\Gamma\qty(\frac{7}{2}+\nu)\Gamma\qty(\frac{7}{2}-\nu){_2}F_1\qty(\frac{7}{2}-\nu,\frac{7}{2}+\nu,4;1+\frac{y}{2}).
        \end{split}
    \end{align}\\
\end{subequations}

\noindent The stochastic approach relies on quantum fluctuations being stretched to classical perturbations by the expansion of the Universe. This approximation is only valid for large spacetime separations and therefore we will only be interested in the leading order term in a large $y$ expansion. We can expand the quantum field correlator to give\\

\begin{equation}
    \label{asymptotic_sum_QFT_corr}
    \begin{split}
    i\Delta(t,t',\mathbf{x},\mathbf{x}')=&\frac{H^2}{16\pi^2}\Bigg[\frac{\Gamma(-2\nu)\Gamma(1+2\nu)}{\Gamma\qty(\frac{1}{2}+\nu)\Gamma\qty(\frac{1}{2}-\nu)}\sum_{s=0}^{\infty}\frac{\Gamma\qty(\frac{3}{2}+\nu+s)\Gamma\qty(\frac{1}{2}+\nu+s)}{\Gamma\qty(1+2\nu+s)s!}\qty(-\frac{y}{2})^{-\frac{3}{2}-\nu-s}\\&\\
    &+\frac{\Gamma(2\nu)\Gamma(1-2\nu)}{\Gamma\qty(\frac{1}{2}+\nu)\Gamma\qty(\frac{1}{2}-\nu)}\sum_{s=0}^{\infty}\frac{\Gamma\qty(\frac{3}{2}-\nu+s)\Gamma\qty(\frac{1}{2}-\nu+s)}{\Gamma\qty(1-2\nu+s)s!}\qty(-\frac{y}{2})^{-\frac{3}{2}+\nu-s}\Bigg].
    \end{split}
\end{equation}\\

\noindent Thus, we find that taking the leading order terms in the two sums gives\\

\begin{equation}
    \label{qft_field_propagator_leading_order}
    \begin{split}
    i\Delta(t,t',\mathbf{x},\mathbf{x}')=\frac{H^2}{16\pi^2}\Bigg[\frac{\Gamma(2\nu)\Gamma(\frac{3}{2}-\nu)}{\Gamma(\frac{1}{2}+\nu)}\qty(-\frac{y}{2})^{-\frac{3}{2}+\nu}+\frac{\Gamma(-2\nu)\Gamma(\frac{3}{2}+\nu)}{\Gamma(\frac{1}{2}-\nu)}\qty(-\frac{y}{2})^{-\frac{3}{2}-\nu}\Bigg].
    \end{split}
\end{equation}\\

\noindent The first term in the above expression is the leading term in the asymptotic expansion, while the second term is subleading. For light scalar fields where $0<\nu<1/2$, this term is not next-to-leading order.\\

The two spacetime regimes that we wish to consider are equal-time and equal-space. The equal-space correlators to leading order in the two sums are\\

\begin{subequations}
\label{equal-space_correlators}
    \begin{align}
    \label{equal-space_phiphi_corr}
    \begin{split}
        \expval{\hat{T}\hat{\phi}(t,\mathbf{x})\hat{\phi}(t',\mathbf{x})}=&\frac{H^2}{16\pi^2}\frac{\Gamma(2\nu)\Gamma\qty(\frac{3}{2}-\nu)}{\Gamma\qty(\frac{1}{2}+\nu)}\qty(-\frac{e^{H(t-t')}}{4})^{-\frac{3}{2}+\nu}\\&+\frac{H^2}{16\pi^2}\frac{\Gamma(-2\nu)\Gamma\qty(\frac{3}{2}+\nu)}{\Gamma\qty(\frac{1}{2}-\nu)}\qty(-\frac{e^{H(t-t')}}{4})^{-\frac{3}{2}-\nu},\\&
    \end{split}\\
    \label{equal-space_phipi_corr}
    \begin{split}
        \expval{\hat{T}\hat{\phi}(t,\mathbf{x})\hat{\pi}(t',\mathbf{x})}=&\frac{H^3}{16\pi^2}\frac{\alpha\Gamma(2\nu)\Gamma\qty(\frac{3}{2}-\nu)}{\Gamma\qty(\frac{1}{2}+\nu)}\qty(-\frac{e^{H(t-t')}}{4})^{-\frac{3}{2}+\nu}\\&+\frac{H^3}{16\pi^2}\frac{\beta\Gamma(-2\nu)\Gamma\qty(\frac{3}{2}+\nu)}{\Gamma\qty(\frac{1}{2}-\nu)}\qty(-\frac{e^{H(t-t')}}{4})^{-\frac{3}{2}-\nu},\\&
        \end{split}\\
    \label{equal-space_piphi_corr}
    \begin{split}
        \expval{\hat{T}\hat{\pi}(t,\mathbf{x})\hat{\phi}(t',\mathbf{x})}=&-\frac{H^3}{16\pi^2}\frac{\alpha\Gamma(2\nu)\Gamma\qty(\frac{3}{2}-\nu)}{\Gamma\qty(\frac{1}{2}+\nu)}\qty(-\frac{e^{H(t-t')}}{4})^{-\frac{3}{2}+\nu}\\&-\frac{H^3}{16\pi^2}\frac{\beta\Gamma(-2\nu)\Gamma\qty(\frac{3}{2}+\nu)}{\Gamma\qty(\frac{1}{2}-\nu)}\qty(-\frac{e^{H(t-t')}}{4})^{-\frac{3}{2}-\nu},\\&
        \end{split}\\
    \label{equal-space_pipi_corr}
    \begin{split}
        \expval{\hat{T}\hat{\pi}(t,\mathbf{x})\hat{\pi}(t',\mathbf{x})}=&\frac{H^4}{16\pi^2}\frac{\alpha^2\Gamma(2\nu)\Gamma\qty(\frac{3}{2}-\nu)}{\Gamma\qty(\frac{1}{2}+\nu)}\qty(-\frac{e^{H(t-t')}}{4})^{-\frac{3}{2}+\nu}\\&+\frac{H^4}{16\pi^2}\frac{\beta^2\Gamma(-2\nu)\Gamma\qty(\frac{3}{2}+\nu)}{\Gamma\qty(\frac{1}{2}-\nu)}\qty(-\frac{e^{H(t-t')}}{4})^{-\frac{3}{2}-\nu},\\&
        \end{split}
    \end{align}\\
\end{subequations}

\noindent while for equal times they are given by\\

\begin{subequations}
\label{equal-time_quantum_correlators}
    \begin{align}
    \label{equal-time_phiphi_corr}
    \begin{split}
        \expval{\hat{\phi}(t,\mathbf{x})\hat{\phi}(t,\mathbf{x}')}=&\frac{H^2}{16\pi^2}\frac{\Gamma(2\nu)\Gamma\qty(\frac{3}{2}-\nu)}{\Gamma\qty(\frac{1}{2}+\nu)}\qty(\frac{He^{Ht}\abs{\mathbf{x}-\mathbf{x}'}}{2})^{-3+2\nu}\\&+\frac{H^2}{16\pi^2}\frac{\Gamma(-2\nu)\Gamma\qty(\frac{3}{2}+\nu)}{\Gamma\qty(\frac{1}{2}-\nu)}\qty(\frac{He^{Ht}\abs{\mathbf{x}-\mathbf{x}'}}{2})^{-3-2\nu},\\&
    \end{split}\\
    \label{equal-time_phipi_corr}
    \begin{split}
        \expval{\hat{\phi}(t,\mathbf{x})\hat{\pi}(t,\mathbf{x}')}=&-\frac{H^3}{16\pi^2}\frac{\alpha\Gamma(2\nu)\Gamma\qty(\frac{3}{2}-\nu)}{\Gamma\qty(\frac{1}{2}+\nu)}\qty(\frac{He^{Ht}\abs{\mathbf{x}-\mathbf{x}'}}{2})^{-3+2\nu}\\&-\frac{H^3}{16\pi^2}\frac{\beta\Gamma(-2\nu)\Gamma\qty(\frac{3}{2}+\nu)}{\Gamma\qty(\frac{1}{2}-\nu)}\qty(\frac{He^{Ht}\abs{\mathbf{x}-\mathbf{x}'}}{2})^{-3-2\nu},\\&
    \end{split}\\
    \label{equal-time_piphi_corr}
    \begin{split}
        \expval{\hat{\pi}(t,\mathbf{x})\hat{\phi}(t,\mathbf{x}')}=&-\frac{H^3}{16\pi^2}\frac{\alpha\Gamma(2\nu)\Gamma\qty(\frac{3}{2}-\nu)}{\Gamma\qty(\frac{1}{2}+\nu)}\qty(\frac{He^{Ht}\abs{\mathbf{x}-\mathbf{x}'}}{2})^{-3+2\nu}\\&-\frac{H^3}{16\pi^2}\frac{\beta\Gamma(-2\nu)\Gamma\qty(\frac{3}{2}+\nu)}{\Gamma\qty(\frac{1}{2}-\nu)}\qty(\frac{He^{Ht}\abs{\mathbf{x}-\mathbf{x}'}}{2})^{-3-2\nu},\\&
    \end{split}\\
    \label{equal-time_pipi_corr}
    \begin{split}
        \expval{\hat{\pi}(t,\mathbf{x})\hat{\pi}(t,\mathbf{x}')}=&\frac{H^4}{16\pi^2}\frac{\alpha^2\Gamma(2\nu)\Gamma\qty(\frac{3}{2}-\nu)}{\Gamma\qty(\frac{1}{2}+\nu)}\qty(\frac{He^{Ht}\abs{\mathbf{x}-\mathbf{x}'}}{2})^{-3+2\nu}\\&+\frac{H^4}{16\pi^2}\frac{\beta^2\Gamma(-2\nu)\Gamma\qty(\frac{3}{2}+\nu)}{\Gamma\qty(\frac{1}{2}-\nu)}\qty(\frac{He^{Ht}\abs{\mathbf{x}-\mathbf{x}'}}{2})^{-3-2\nu}.
    \end{split}
    \end{align}\\
\end{subequations}

\noindent We note that it is unneccesary to specify the time ordering of the operators for the equal-time correlators since they commute. Therefore, in contrast to timelike separation, there is no ambiguity in the definition of spacelike QFT correlators.\\

\subsection{Comparison between the cut-off stochastic and quantum correlators}
\label{subsec:stoch_v_quantum}

We are now in a position to consider the validity of the cut-off approach with respect to the QFT. We will focus on the equal-time field correlator in this comparison. By substituting the cut-off noise amplitudes of Eq. (\ref{noise_amplitudes_modes}) into Eq. (\ref{sigma_q,p-->sigma_phi,pi}), we can express the $(q,p)$ noise amplitudes in terms of Hankel functions and hence the stochastic field correlator is given by\\

\begin{equation}
    \label{stochastic_field_mode_correlator}
    \begin{split}
        \expval{\phi(0,\mathbf{0})\phi(0,\mathbf{x})}_{cut}=&\frac{H^2\epsilon^3}{256\nu^2\pi\alpha}\abs{\epsilon\mathcal{H}_{\nu-1}^{(1)}(\epsilon)-2\nu\mathcal{H}_{\nu}^{(1)}(\epsilon)-\epsilon\mathcal{H}_{\nu+1}^{(1)}(\epsilon)}^2\abs{H\mathbf{x}}^{-3+2\nu}
        \\&
        +\frac{H^2\epsilon^3}{256\nu^2\pi\beta}\abs{\epsilon\mathcal{H}_{\nu-1}^{(1)}(\epsilon)+2\nu\mathcal{H}_{\nu}^{(1)}(\epsilon)-\epsilon\mathcal{H}_{\nu+1}^{(1)}(\epsilon)}^2\abs{H\mathbf{x}}^{-3-2\nu}.
    \end{split}
\end{equation}\\

\noindent Note that we exclude the $\abs{Ha(t)\mathbf{x}}^{-3}$ term in what follows as that is an additional stochastic term that does not appear in the quantum correlator. Comparing this expression with Eq. (\ref{qft_field_propagator_leading_order}), we find that the cut-off approach will reproduce the leading order term of the quantum correlator if\\

\begin{equation}
    \label{SY=quantum_alpha}
    \begin{split}
    \frac{H^2\epsilon^3}{256\nu^2\pi\alpha}\abs{\epsilon\mathcal{H}_{\nu-1}^{(1)}(\epsilon)-2\nu\mathcal{H}_{\nu}^{(1)}(\epsilon)-\epsilon\mathcal{H}_{\nu+1}^{(1)}(\epsilon)}^2=\frac{H^2}{16\pi^2}\Bigg[\frac{\Gamma(\frac{3}{2}-\nu)\Gamma(2\nu)4^{\frac{3}{2}-\nu}}{\Gamma(\frac{1}{2}+\nu)}\Bigg].
    \end{split}
    \end{equation}\\

\noindent This can only be true if $\epsilon$ is mass-dependent. Note that this solution will not reproduce the leading term in the second sum of Eq. (\ref{asymptotic_sum_QFT_corr}).\\

To make this comparison between the cut-off stochastic and quantum correlators more concrete we define a quantum noise as\\

\begin{equation}
    \label{quantum_noise_contribution}
    \frac{H^2}{16\pi^2}\frac{\Gamma(\frac{3}{2}-\nu)\Gamma(2\nu)4^{\frac{3}{2}-\nu}}{\Gamma(\frac{1}{2}+\nu)}=\frac{1}{1-\frac{\alpha}{\beta}}\frac{\sigma_{Q,qq}^2}{2H\alpha}
\end{equation}\\

\begin{figure}[t]
    \centering
    \includegraphics[width=120mm]{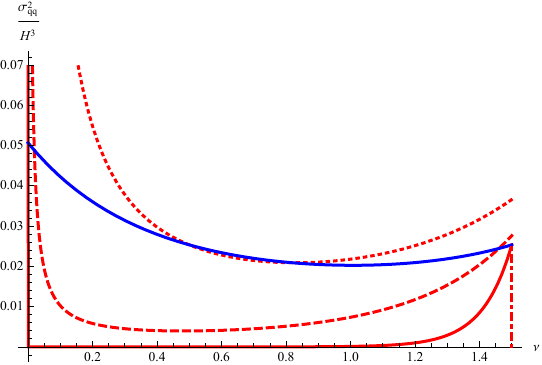}
    \caption{The quantum (blue) $\sigma_{Q,qq}^2$ and the cut-off (red) noises $\sigma_{cut,qq}^2$ with $\epsilon=0$ (dot-dashed), $\epsilon=0.01$ (solid), $\epsilon=0.5$ (dashed) and $\epsilon=0.99$ (dotted) is plotted as a function of $\nu$. We see that the two approaches don't agree for any value of $\epsilon$ and therefore the cut-off procedure is unable to reproduce the quantum field correlator for all masses.}
    \label{fig:noise_plot}
\end{figure}

\noindent If the stochastic noise has the amplitude $\sigma_{qq}^2=\sigma_{Q,qq}^2$ then the leading term in the equal-time stochastic field correlator (\ref{spacelike_phi-phi_stochastic_correlator}) is equal to the leading term in the equal-time quantum field correlator (\ref{equal-time_phiphi_corr}). To compare this with the cut-off method, we plot $\sigma_{Q,qq}^2$ and $\sigma_{cut,qq}^2$, for various values of $\epsilon$, as a function of $\nu$ in Fig. \ref{fig:noise_plot}. We observe that the cut-off procedure correctly calculates the leading-order amplitude in the small mass $m<<H$ limit as $\epsilon\longrightarrow0$ but beyond this the results diverge from each other.\\

\subsection{The matching procedure}
\label{subsec:matching}

As an alternative to the cut-off procedure, we propose to generalise the stochastic approach. We assume the stochastic equations are that of Eq. (\ref{Langevin_equation0}) but we keep the noise general, as we do throughout Sections \ref{subsec:fokker-planck} and \ref{subsec:correlators}. Then, the procedure for calculating stochastic correlators remains unchanged, but we match the noise amplitude with the equal-time quantum correlators (\ref{equal-time_quantum_correlators}). This not only trivially reproduces the 2-point equal-time correlators to leading order, but also reproduces all higher-order equal-time correlators correctly for the matched noise.\\

The noise matrix has 3 degrees of freedom: $\sigma_{qq}^2$, $\sigma_{qp}^2$ and $\sigma_{pp}^2$. The primary constraint is that the leading order behaviour of the equal-time quantum correlators is reproduced by their stochastic counterparts; in other words, it satisfies Eq. (\ref{quantum_noise_contribution}). We also choose to match with the subleading contribution, which results in a constraint on $\sigma_{pp}^2$. Finally, we constrain $\sigma_{qp}^2$ such that the equal-space quantum correlators are also reproduced by their stochastic counterparts up to a complex phase. This amounts to the continuation from equal space to equal time via $e^{H(t-t')}\longrightarrow(Ha(t)\abs{\mathbf{x}-\mathbf{x}'})^2$. We suggest that this complex phase should not feature in the stochastic approach since the stochastic correlators are real by definition. This may be an indication of long-distance quantum features that are not accounted for, though more work is required to understand this more concretely.\\

With these constraints in place, our matched $(q,p)$ noise contributions are given by\\

\begin{subequations}
    \label{matched_(p,q)_noise}
\begin{align}
        \label{matched_sigma_qq}
        \sigma_{Q,qq}^2&=\frac{H^3\alpha\nu}{4\pi^2\beta}\frac{\Gamma(2\nu)\Gamma(\frac{3}{2}-\nu)4^{\frac{3}{2}-\nu}}{\Gamma(\frac{1}{2}+\nu)},\\
        \label{matched_sigma_qp}
        \sigma_{Q,qp}^2&=0,\\
        \label{matched_sigma_p}
        \sigma_{Q,pp}^2&=\frac{H^5\beta^2\nu}{4\pi^2}\frac{\Gamma(-2\nu)\Gamma(\frac{3}{2}+\nu)4^{\frac{3}{2}+\nu}}{\Gamma(\frac{1}{2}-\nu)}.
\end{align}\\
\end{subequations}

\noindent These can then be substituted into the equal-time stochastic correlators (\ref{equal-time_(phi,pi)_stochastic_correlators}) to give (\ref{equal-time_quantum_correlators}). Note that this choice simplifies the Fokker-Planck equation (\ref{p,q_fokker-planck}) to two uncorrelated 1-dimensional Fokker-Planck equations for $q$ and $p$. Thus, the separation of the time-evolution operator discussed in Sec. \ref{subsec:correlators} applies to all correlators.\\

We note that the matching stochastic approach reproduces the equal-time quantum correlators for all masses, including when $\nu$ becomes imaginary. We suspect that this is only true for free fields and that, for self-interacting theories, the stochastic approach will break down if the fields are too heavy since the expansion of the Universe will no longer be sufficient to promote quantum fluctuations to classical perturbations. More work is required to understand this in more detail. However, the purpose of the stochastic approach is to replace the IR divergent perturbation theory. This is only necessary for light scalar fields, where we expect the stochastic approach to work.\\

\subsection{Quantum variance}
\label{subsec:variance}

Before considering higher-order correlators, we compare the matched stochastic and quantum variances. We have calculated the general form of the stochastic variances in Section \ref{subsubsec:stoch_variance}. Substituting the matched noise of Eq. (\ref{matched_(p,q)_noise}) into Eq. (\ref{stoch_variance_noise}), we find that\\

\begin{subequations}
    \label{matched_stoch_variances}
    \begin{align}
        \label{matched_phiphi_variance}
        \expval{\phi^2}&=\frac{H^2}{16\pi^2}\qty[\frac{\Gamma(\frac{3}{2}-\nu)\Gamma(2\nu)4^{\frac{3}{2}-\nu}}{\Gamma(\frac{1}{2}+\nu)}+\frac{\Gamma(\frac{3}{2}+\nu)\Gamma(-2\nu)4^{\frac{3}{2}+\nu}}{\Gamma(\frac{1}{2}-\nu)}],\\
        \label{matched_phipi_variance}
        \expval{\phi\pi}&=-\frac{H^2}{16\pi^2}\qty[\frac{\alpha\Gamma(\frac{3}{2}-\nu)\Gamma(2\nu)4^{\frac{3}{2}-\nu}}{\Gamma(\frac{1}{2}+\nu)}+\frac{\beta\Gamma(\frac{3}{2}+\nu)\Gamma(-2\nu)4^{\frac{3}{2}+\nu}}{\Gamma(\frac{1}{2}-\nu)}],\\
        \label{matched_pipi_variance}
        \expval{\pi^2}&=\frac{H^4}{16\pi^2}\qty[\frac{\alpha^2\Gamma(\frac{3}{2}-\nu)\Gamma(2\nu)4^{\frac{3}{2}-\nu}}{\Gamma(\frac{1}{2}+\nu)}+\frac{\beta^2\Gamma(\frac{3}{2}+\nu)\Gamma(-2\nu)4^{\frac{3}{2}+\nu}}{\Gamma(\frac{1}{2}-\nu)}].
    \end{align}
\end{subequations}

\noindent In the quantum theory, even for free fields, these quantities contain ultraviolet divergences since they are governed by short-distance dynamics. It is therefore not appropriate to compare this quantum object with its stochastic equivalent because the stochastic approach is only valid for large spacetime separations. We only include it here to allow us to compare higher-order correlators. Its inclusion in these expressions allows us to isolate the ambiguity surrounding the UV divergences.\\

To see this more clearly, we include the quantum variances, which are given by \cite{Birrell-Davies_book,tagirov_QFT_deSitter,bunch_davies_1978}\\

\begin{subequations}
    \label{quantum_variance}
    \begin{align}
    \label{quantum_phiphi_variance}
    \expval{\hat{\phi}^2}&=\frac{H^2}{16\pi^2}\Gamma\qty(\frac{3}{2}-\nu)\Gamma\qty(\frac{3}{2}+\nu){_2}F_1\qty(\frac{3}{2}+\nu,\frac{3}{2}-\nu,2;1),\\
    \label{quantum_phipi_variance}
    \expval{\hat{\phi}\hat{\pi}}&=0,\\
    \label{quantum_pipi_variance}
    \expval{\hat{\pi}^2}&=-\frac{\alpha\beta H^4}{64\pi^2}\Gamma\qty(\frac{3}{2}-\nu)\Gamma\qty(\frac{3}{2}+\nu){_2}F_1\qty(\frac{7}{2}+\nu,\frac{7}{2}-\nu,4;1).
    \end{align}\\
\end{subequations}

\noindent It is clear that the variances do not match between quantum and stochastic approaches, which is to be expected as discussed. In the small mass $m<<H$ limit , the variances are given by\\

\begin{subequations}
\label{massless_variance}
\begin{align}
        \label{massless_field_variance}
        \expval{\hat{\phi}^2}\eval_{m<<H}&=\frac{3H^4}{8\pi^2m^2}, & \expval{\phi^2}\eval_{m<<H}&=\frac{3H^4}{8\pi^2m^2};\\
        \label{massless_phipi_variance}
        \expval{\hat{\phi}\hat{\pi}}\eval_{m<<H}&=0, &  \expval{\phi\pi}\eval_{m<<H}&=-\frac{17H^3}{8\pi^2};\\
        \label{massless_pipi_variance}
        \expval{\hat{\pi}^2}\eval_{m<<H}&=-\frac{3H^4}{32\pi^2}, &  \expval{\pi^2}\eval_{m<<H}&=\frac{6H^2}{\pi^2}.
\end{align}\\
\end{subequations}

\noindent One can see that the field variances match. This is because the $1/m^2$ term comes from the leading order contribution to the asymptotic expansion, precisely as the stochastic approach reproduces, due to its IR divergence in the small mass limit. For the $\phi-\pi$ and $\pi-\pi$ variances, subleading terms in the asymptotic expansion contribute to the leading terms in the small mass expansion and thus the stochastic and quantum variances don't match even in the small mass limit.\\

\subsection{Higher-order correlators}
\label{subsec:higher-order_correlators}

We are finally in a position to consider higher-order correlators. This is done using the joint unequal-time PDF (\ref{joint_PDF_2-point}) with the matched noise of Eq. (\ref{matched_(p,q)_noise}). Thus, all equal-space higher-order correlators can be calculated via\\

\begin{equation}
    \label{higher-correlators_PDF}
    \begin{split}
    \langle f(\phi(0,\mathbf{x}),\pi(0,\mathbf{x}))&g(\phi(t,\mathbf{x}),\pi(t,\mathbf{x}))\rangle\\&=\int d\phi_0\int d\phi \int d\pi_0\int d\pi P_2(\phi_0,\phi,\pi_0,\pi) f_0(\phi_0,\pi_0) g(\phi,\pi).
    \end{split}
\end{equation}\\

\noindent The PDF of Eq. (\ref{joint_PDF_time}) is Gaussian and therefore Wick's theorem can be used to compute correlators of arbitrary functions of $\phi$ and $\pi$. The only quantities that need explicit calculation are the $\phi-\phi$, $\phi-\pi$, $\pi-\phi$ and $\pi-\pi$ correlators with their corresponding variances, all of which have been done. This also means that we have implicitly proven that the stochastic and quantum results agree to leading asymptotic order for all diagrammatic contributions to all correlators. Thus, the matching of the stochastic and quantum 2-point field correlator is not arbitrary but provides a key calculational tool for dealing with correlators in de Sitter.\\

As an example, let's consider the $\phi^2$ correlator. Using Wick's theorem, one can calculate the stochastic correlator as\\

\begin{equation}
    \label{phi^2_correlator}
    \expval{\phi(0,\mathbf{0})^2\phi(0,\mathbf{x})^2}=\expval{\phi^2}^2+2\qty(\expval{\phi(0)\phi(\mathbf{x}})^2.
\end{equation}\\

Alternatively, one can calculate the $\phi^2$ stochastic correlator using the methods outlined in Sec. \ref{subsec:correlators}. With $\sigma_{qp}^2=0$, it is possible to write\\

\begin{equation}
    \label{time-ev_operator_qp=0}
    U(q_0,q,p_0,p;t)=U_q(q_0,q;t)U_p(p_0,p;t),
\end{equation}\\

\noindent and hence all timelike $(q,p)$-correlators can be calculated from\\

\begin{equation}
    \label{<fg>}
    \expval{f(q,p;0)g(q,p;t)}=\int dq_0\int dq\int dp_0\int dp P_{eq}(q_0,p_0)U_q(q_0,q;t)U_p(p_0,p;t)f(q_0,p_0)g(q,p).
\end{equation}\\

\noindent Then, all higher-order $(\phi,\pi)$-correlators can be related to this via Eq. (\ref{p,q-->pi,phi}) and hence calculated. For the equal-space $\phi^2$-correlator\\

\begin{equation}
    \label{<phi^2phi^2>}
    \expval{\phi(0,\mathbf{x})^2\phi(t,\mathbf{x})^2}=\frac{1}{(1-\frac{\alpha}{\beta})^2}\expval{\qty(-\frac{1}{\beta H}p(0)+q(0))^2\qty(-\frac{1}{\beta H}p(t)+q(t)^2)}.
\end{equation}\\

\noindent By expanding the brackets, we see that this correlator will depend on 9 $(q,p)$-correlators. To ensure we get all leading-order contributions for this correlator, we perform the spectral expansion with a sum up to $n=2$ for the time-evolution operators of Eq. (\ref{time-ev_solution}). Evaluating the $(q,p)$-correlators, we find that many of them are zero. The remaining non-zero correlators are\\

\begin{subequations}
    \label{non-zero_pq_corr}
    \begin{equation}
        \label{<p02p2>}
        \expval{p(0)^2p(t)^2}=\frac{\sigma_{Q,pp}^4}{4H^2\beta^2}\qty(1+2e^{-2\beta H t});
    \end{equation}
    \begin{equation}
        \label{p02q2}
        \expval{p(0)^2q(t)^2}=\frac{\sigma_{Q,pp}^2\sigma_{Q,qq}^2}{4H^2\alpha\beta};
    \end{equation}
    \begin{equation}
        \label{p0q0pq}
        \expval{p(0)q(0)p(t)q(t)}=\frac{\sigma_{Q,pp}^2\sigma_{Q,qq}^2}{4H^2\alpha\beta}e^{-3 Ht};
    \end{equation}
    \begin{equation}
        \label{q02p2}
        \expval{q(0)^2p(t)^2}=\frac{\sigma_{Q,pp}^2\sigma_{Q,qq}^2}{4H^2\alpha\beta};
    \end{equation}
    \begin{equation}
        \label{q02q2}
        \expval{q(0)^2q(t)^2}=\frac{\sigma_{Q,qq}^4}{4H^2\alpha^2}\qty(1+2e^{-\alpha Ht}).
    \end{equation}\\
\end{subequations}

\noindent Since we have chosen the noise such that timelike and spacelike correlators are related by $e^{H(t-t')}\longrightarrow(Ha(t)\abs{\mathbf{x}-\mathbf{x}'})^2$, the equal-time $\phi^2$ stochastic correlator is given by\\

\begin{equation}
    \label{phi2_stochastic_correlator}
    \begin{split}
    \expval{\phi(t,\mathbf{0})^2\phi(t,\mathbf{x})^2}=&\qty(\frac{1}{1-\frac{\alpha}{\beta}})^2\Bigg[\qty(\frac{\sigma_{Q,qq}^2}{2 H\alpha}+\frac{\sigma_{Q,pp}^2}{2 H^3\beta^3})^2\\&+2\qty(\frac{\sigma_{Q,qq}^2}{2 H\alpha}\abs{Ha(t)\mathbf{x}}^{-2\alpha}+\frac{\sigma_{Q,pp}^2}{2 H^3\beta^3}\abs{Ha(t)\mathbf{x}}^{-2\beta})^2 \Bigg]       
    \end{split}
\end{equation}\\

\noindent The first term is the square of the stochastic field variance, Eq. (\ref{stoch_phiphi_variance_noise}), while the second term is proportional to the square of the stochastic 2-point field correlator, Eq. (\ref{spacelike_phi-phi_stochastic_correlator}). Hence, the spacelike $\phi^2$ correlator is given by Eq. (\ref{phi^2_correlator}), as expected by Wick's theorem.\\

Wick's theorem can also be used to evaluate the quantum $\phi^2$-correlator to give\\

\begin{equation}
    \label{quantum_phi2_correlator}
    \expval{\hat{\phi}(t,\mathbf{0})^2\hat{\phi}(t,\mathbf{x})^2}=\expval{\hat{\phi}^2}^2+2\qty(i\Delta(\mathbf{x}))^2.
\end{equation}\\

\noindent This is equal in form to its stochastic counterpart, with the obvious discrepancy between the variance.\\

\section{Discussion}
\label{sec:discussion}

\subsection{The small mass $m<<H$ limit}
\label{subsec:comparison_work}

The new matching technique of evaluating noise has proven to be effective in calculating the correlators at all orders in mass. It is now instructive to compare this matched noise with the cut-off method. In the small mass limit, the matched and cut-off noises are\\

\begin{subequations}
\begin{align}
\label{massless_noise_comparison}
        \sigma_{Q,qq}^2\eval_{m<<H}&=\frac{H^3}{4\pi^2}, &   \sigma_{cut,qq}^2\eval_{m<<H}&=\frac{H^3}{4\pi^2}(1+\frac{\epsilon^2}{3}+\frac{\epsilon^4}{9});\\
        \label{qp_massless_noise_comparison}
        \sigma_{Q,qp}^2\eval_{m<<H}&=0, & \sigma_{cut,qp}^2\eval_{m<<H}&=\frac{H^4}{4\pi^2}\qty(-\epsilon^2+\frac{\epsilon^4}{3});\\
        \sigma_{Q,pp}^2\eval_{m<<H}&=\frac{36H^5}{\pi^2}, & \sigma_{cut,pp}^2\eval_{m<<H}&=\frac{H^5}{4\pi^2}\epsilon^4.
\end{align}\\
\end{subequations}

One can see that the cut-off and matched $qq$ and $qp$ noises are equal in the limit $\epsilon\longrightarrow0$, the former as indicated by Fig. \ref{fig:noise_plot}. However, the $pp$ noises are not. This suggests that the cut-off method is only capable of reproducing the leading order behaviour, even in the small mass limit, whereas the matching procedure reproduces both leading and subleading. We can make this statement more precise by writing the field correlator for both the cut-off and matching procedures in the small mass limit. This is given by\\

\begin{subequations}
\label{massless_field_corr}
\begin{align}
    \label{massless_ext_SY_field_corr}
    \expval{\phi(t,\mathbf{0})\phi(t,\mathbf{x})}_{cut}\eval_{\epsilon=0}&=\frac{3H^4}{8\pi^2m^2}\abs{Ha(t)\mathbf{x}}^{-\frac{2m^2}{3H^2}}\\
    \label{massless_matched_field_corr}
    \expval{\phi(t,\mathbf{0})\phi(t,\mathbf{x})}_{M}&=\frac{3H^4}{8\pi^2m^2}\abs{Ha(t)\mathbf{x}}^{-\frac{2m^2}{3H^2}}+\frac{2H^2}{3\pi^2}\abs{Ha(t)\mathbf{x}}^{-6+\frac{2m^2}{3H^2}}
\end{align}\\
\end{subequations}

\noindent We can see that both methods produce the leading order behaviour but only the matching method reproduces the subleading contribution. The subleading contribution is negligible compared to the mass-divergent leading term, so the cut-off method is sufficient in this limit, as expected. The leading term is also precisely the result of Starobinsky and Yokoyama's original work in the overdamped limit.\\ 

\subsection{Concluding Remarks}

The stochastic approach can be applied to free scalar fields beyond the slow roll limit. If the noise amplitude in the stochastic equations is correct, equal-time correlators agree with the QFT results at asymptotically long distances, which is the relevant regime for cosmological observations. Timelike correlators are also reproduced up to an overall complex phase.\\

However, the cut-off procedure - a naive extension of Starobinsky and Yokoyama’s original approach – only gives the correct noise term if the cut-off $\epsilon$ depends on the mass in a specific way. We determine this mass-dependence by matching with the QFT. Alternatively, one can obtain the noise term directly by matching with the QFT, without introducing a cut-off. Both these procedures are equivalent and both require knowledge of the asymptotic QFT correlators.\\

Regardless of how one chooses to evaluate the noise, the key result is that the stochastic approach is a viable alternative to QFT for free fields if the noise contribution is given by Eq. (\ref{matched_sigma_qq}). This motivates future work, where this method will be applied to interacting theories, as it suggests that the stochastic approach offers the ability to calculate objects that cannot be found by standard QFT procedures. Because the exact quantum correlators are not known in that case, one would ideally need a way of determining the noise amplitude directly from the microscopic picture without matching. Currently it is not known how to do that, but by showing what the answer needs to be in the free case, our results are an important step in that direction.

\acknowledgements A.C. was supported by a UK Science and Technology Facility Council studentship. A.R. was supported by STFC grants ST/P000762/1 and ST/T000791/1 and IPPP Associateship. We would like to thank Grigoris Pavliotis and Lucas Pinol for useful discussions.

\appendix

\section{Derivation of the eigenspectrum}
\label{app:eigenspectrum}

In this appendix, we derive Eq. (\ref{time-ev_solution}), the expression for the time-evolution operators, via a spectral expansion \cite{Markkanen_2019}. We will explicitly derive $U_q(q_0,q;t)$ here; the derivation of $U_p(p_0,p;t)$ follows in precisely the same way. We begin with Eq. (\ref{q_time-ev_pde})\\

\begin{equation}
    \label{q_time-ev_pde_appendix}
    \pdv{U_q(q_0,q;t)}{t}=\alpha H U_q(q_0,q;t)+\alpha H q \pdv{U_q(q_0,q;t)}{q}+\frac{1}{2}\sigma_{qq}^2\pdv[2]{U_q(q_0,q;t)}{q}.
\end{equation}\\

\noindent which holds for any $q_0$. We define\\

\begin{equation}
    \label{redef_time-ev}
    U_q(q_0,q;t)=e^{-\frac{\alpha H}{2\sigma_{qq}^2}(q^2-q_0^2)}\Tilde{U}_q(q_0,q;t)
\end{equation}\\

\noindent such that Eq. (\ref{q_time-ev_pde_appendix}) becomes\\

\begin{equation}
    \label{redef_time-ev_pde}
    \pdv{\Tilde{U}_q(q_0,q;t)}{t}=\frac{1}{2}\sigma_{qq}^2\pdv[2]{\Tilde{U}_q(q_0,q;t)}{q}+\frac{1}{2}\qty(H\alpha-\frac{H^2\alpha^2}{\sigma_{qq}^2}q^2)\Tilde{U}_q(q_0,q;t).
\end{equation}\\

\noindent which is solved to give\\

\begin{equation}
    \label{redef_time-ev_n}
    \Tilde{U}_q(q_0,q;t)=\sum_n e^{-n \alpha H t}Q_n(q)Q_n(q_0)
\end{equation}\\

\noindent where\\

\begin{equation}
    \label{eigenequation}
    0=\pdv[2]{Q_n(q)}{q}+\qty(\frac{2H \alpha}{\sigma_{qq}^2}\qty(n+\frac{1}{2})-\frac{H^2\alpha^2}{\sigma_{qq}^4}q^2)Q_n(q).
\end{equation}\\

\noindent This equation is recognised as that of a quantum harmonic oscillator. Thus, these eigenstates can be found through standard procedures, resulting in Eq. (\ref{eigenfunctions}), and so the time-evolution operator is given by Eq. (\ref{q_time-ev_solution}), as required. We note that since the right-hand side of Eq. (\ref{eigenequation}) is a self-adjoint operator, the eigenfunctions obey the orthonormality and completeness relations\\

\begin{subequations}
    \begin{equation}
        \label{orthonormality_condition}
        \int dq Q_n(q)Q_m(q)=\delta_{nm},
    \end{equation}
    \begin{equation}
        \label{completeness}
        \sum_n Q_n(q)Q_n(q')=\delta(q-q').
    \end{equation}\\
\end{subequations}

\bibliographystyle{unsrt}
\bibliography{stochfree.bib}
\end{document}